\documentclass[12pt,prd,showpacs,tightenlines,nofootinbib]{revtex4}
\usepackage{bm}
\usepackage{graphics}
\usepackage{rotating}
\usepackage{epsfig}
\usepackage{longtable}
\begin{document}
\begin{flushright}{HU-EP-09/14}\end{flushright}
\title{
Mass spectra and Regge trajectories
of light mesons in the relativistic quark model}
\author{D. Ebert$^{1}$, R. N. Faustov$^{2}$  and V. O. Galkin$^{2}$}
\affiliation{
$^1$ Institut f\"ur Physik, Humboldt--Universit\"at zu Berlin,
Newtonstr. 15, D-12489  Berlin, Germany\\
$^2$ Dorodnicyn Computing Centre, Russian Academy of Sciences,
  Vavilov Str. 40, 119991 Moscow, Russia}

\begin{abstract}
Masses of the ground, orbitally and radially excited  states of quark-antiquark
mesons composed from the light ($u,d,s$) quarks are calculated within
the framework of the relativistic quark model based on the
quasipotential approach. The relativistic treatment of the light quark
dynamics results in mass spectra which agree well with available
experimental data for the masses of the most well-established states. 
The Regge trajectories for angular and radial excitations are constructed,
and their linearity, parallelism and equidistance are verified. 
The assignment of experimentally observed light mesons to
particular Regge trajectories is based on their masses and
quantum numbers. 
\end{abstract}


\maketitle

\section{Introduction}
\label{sec:intr}

Last years an extensive analysis of the data on highly excited light
non-strange meson states up to a mass of 2400 MeV collected by the
Crystal Barrel experiment at LEAR (CERN)  has been published
\cite{bugg,Anisovich}. 
The classification of these new data requires a better
theoretical understanding of light meson mass spectra.
The aim of this paper is to apply the relativistic quark model which
proved to be successful in studying various properties of heavy
hadrons to the calculation of the masses of the radially and orbitally excited
light meson states.  All main
assumptions and fixed values of model parameters are preserved in the
present investigation. Light quarks are treated fully relativistically
without the $v/c$ expansion. Various non-strange and strange meson
states with masses up to 2500 MeV are considered.  
 This is especially
important, since light exotic states (such as tetraquarks, glueballs,
hybrids)  predicted by quantum chromodynamics (QCD) are expected to have masses
in this range \cite{at,kz,glb,achasov}. 
The experimental data
show that a large degeneracy emerges in the spectra 
of the orbitally and radially excited resonances.  It is argued \cite{glozman}
that the states of the same spin with different isospins and opposite
parities are approximately degenerate in the interval 1700-2400
MeV. An intensive debate is going on now in the literature 
whether the chiral symmetry is restored for
highly excited states (see e.g. \cite{glozman,shifvain} and references
therein). Various phenomenological and theoretical arguments, such
as quasiclassical considerations, AdS/QCD etc. are used.

A vast literature on the light meson spectroscopy is
available. Different attempts to study light mesons on the basis of
the relativized quark model \cite{gi}, the
Dyson-Schwinger and Bethe-Salpeter equations \cite{mr,k,bnps}, the
Tamm-Dancoff method \cite{lse},  chiral quark
models with spontaneous symmetry breaking like
the Nambu-Jona-Lasinio model \cite{erv}, 
finite-energy sum rules in QCD \cite{kp},
lattice QCD \cite{ak}, AdS/QCD models \cite{fork,af}, etc. were
undertaken. Therefore we mostly refer to the recent 
reviews where the references to earlier review and original papers
can be found. 

In Refs.~\cite{lmes,lmesff} we studied the masses of ground and
radially excited states of light mesons  on the
basis of the three-dimensional relativistic wave equation with the
QCD-motivated potential. In this analysis we took into account the highly
relativistic dynamics of light quarks and carried out
all calculations
without either the $v/c$ or $1/m_q$ expansions. We also used the
expression for the QCD coupling constant $\alpha_s$ which exhibits
freezing at small values of the momentum transfer. Good overall
agreement of the obtained 
predictions and experimental data was found. The consistent
relativistic treatment of the light quark dynamics resulted in a
nonlinear dependence of the bound state equation on the meson mass
which allowed to get the correct values of the pion and kaon masses in
the model with the explicitly broken chiral symmetry. The obtained
wave functions of 
the pion and kaon were successfully applied for the relativistic
calculation of their decay constants and electromagnetic form factors
\cite{lmesff}. 
Recently, in the framework of the same approach we
calculated masses of the ground-state light  
tetraquarks using the diquark-antidiquark picture \cite{ltetr}. It was
found that scalar mesons with masses below 1 GeV agree well with the
light-tetraquark interpretation. Indeed, it explains naturally the
peculiar inverted pattern of the mass ordering of the lightest scalar
flavour $SU(3)$ nonet. Here we
investigate the Regge trajectories both in ($M^2$,$J$) and ($M^2$,
$n_r$) planes ($M$ is the mass, $J$ is the spin and
 $n_r$ is the radial quantum number of the meson state), check their
 linearity and equidistance which follow from experimental data
 \cite{anisovich}.   
The assignment of experimentally observed mesons to particular Regge
trajectories is proposed.

The paper is organized as follows. In Section~\ref{sec:rqm} we describe 
the relativistic quark model, giving its main assumptions and 
parameters which were fixed in previous considerations. The relativistic
quasipotential of the light quark-antiquark interaction in the
meson is
constructed in Sec~\ref{sec:qpot}. The procedure which makes this
potential local  and avoids the arising fictitious singularities  
is described in detail. In Sec.~\ref{sec:rd} the obtained results 
for the masses of orbital and radial excitations of light mesons are 
presented and compared with available experimental data. 
Finally, the Regge trajectories are constructed.
Their linearity, parallelism and equidistance is verified, and
the slopes of different trajectories are compared.  
Section~\ref{sec:concl} contains our conclusions.        

\section{Relativistic quark model}
\label{sec:rqm}

  In the relativistic quark model based on the quasipotential approach
  a meson is described by the wave 
function of the bound quark-antiquark state, which satisfies the
quasipotential equation  of the Schr\"odinger type~\cite{efg}
\begin{equation}
\label{quas}
{\left(\frac{b^2(M)}{2\mu_{R}}-\frac{{\bf
p}^2}{2\mu_{R}}\right)\Psi_{M}({\bf p})} =\int\frac{d^3 q}{(2\pi)^3}
 V({\bf p,q};M)\Psi_{M}({\bf q}),
\end{equation}
where the relativistic reduced mass is
\begin{equation}
\mu_{R}=\frac{E_1E_2}{E_1+E_2}=\frac{M^4-(m^2_1-m^2_2)^2}{4M^3},
\end{equation}
and $E_1$, $E_2$ are given by
\begin{equation}
\label{ee}
E_1=\frac{M^2-m_2^2+m_1^2}{2M}, \quad E_2=\frac{M^2-m_1^2+m_2^2}{2M}.
\end{equation}
Here $M=E_1+E_2$ is the meson mass, $m_{1,2}$ are the quark masses,
and ${\bf p}$ is their relative momentum.  
In the center-of-mass system the relative momentum squared on mass shell 
reads
\begin{equation}
{b^2(M) }
=\frac{[M^2-(m_1+m_2)^2][M^2-(m_1-m_2)^2]}{4M^2}.
\end{equation}

The kernel 
$V({\bf p,q};M)$ in Eq.~(\ref{quas}) is the quasipotential operator of
the quark-antiquark interaction. It is constructed with the help of the
off-mass-shell scattering amplitude, projected onto the positive
energy states. 
Constructing the quasipotential of the quark-antiquark interaction, 
we have assumed that the effective
interaction is the sum of the usual one-gluon exchange term with the mixture
of long-range vector and scalar linear confining potentials, where
the vector confining potential
contains the Pauli interaction. The quasipotential is then defined by
  \begin{equation}
\label{qpot}
V({\bf p,q};M)=\bar{u}_1(p)\bar{u}_2(-p){\mathcal V}({\bf p}, {\bf
q};M)u_1(q)u_2(-q),
\end{equation}
with
$${\mathcal V}({\bf p},{\bf q};M)=\frac{4}{3}\alpha_sD_{ \mu\nu}({\bf
k})\gamma_1^{\mu}\gamma_2^{\nu}
+V^V_{\rm conf}({\bf k})\Gamma_1^{\mu}
\Gamma_{2;\mu}+V^S_{\rm conf}({\bf k}),$$
where $\alpha_s$ is the QCD coupling constant, $D_{\mu\nu}$ is the
gluon propagator in the Coulomb gauge
\begin{equation}
D^{00}({\bf k})=-\frac{4\pi}{{\bf k}^2}, \quad D^{ij}({\bf k})=
-\frac{4\pi}{k^2}\left(\delta^{ij}-\frac{k^ik^j}{{\bf k}^2}\right),
\quad D^{0i}=D^{i0}=0,
\end{equation}
and ${\bf k=p-q}$; $\gamma_{\mu}$ and $u(p)$ are 
the Dirac matrices and spinors
\begin{equation}
\label{spinor}
u^\lambda({p})=\sqrt{\frac{\epsilon(p)+m}{2\epsilon(p)}}
\left(
\begin{array}{c}1\cr {\displaystyle\frac{\bm{\sigma}
      {\bf  p}}{\epsilon(p)+m}}
\end{array}\right)\chi^\lambda,
\end{equation}
with $\epsilon(p)=\sqrt{p^2+m^2}$.
The effective long-range vector vertex is
given by
\begin{equation}
\label{kappa}
\Gamma_{\mu}({\bf k})=\gamma_{\mu}+
\frac{i\kappa}{2m}\sigma_{\mu\nu}k^{\nu},
\end{equation}
where $\kappa$ is the Pauli interaction constant characterizing the
anomalous chromomagnetic moment of quarks. Vector and
scalar confining potentials in the nonrelativistic limit reduce to
\begin{eqnarray}
\label{vlin}
V^V_{\rm conf}(r)&=&(1-\varepsilon)(Ar+B),\nonumber\\ 
V^S_{\rm conf}(r)& =&\varepsilon (Ar+B),
\end{eqnarray}
reproducing 
\begin{equation}
\label{nr}
V_{\rm conf}(r)=V^S_{\rm conf}(r)+V^V_{\rm conf}(r)=Ar+B,
\end{equation}
where $\varepsilon$ is the mixing coefficient. 

All the model parameters have the same values as in our previous
papers \cite{egf,efg}.
The light constituent quark masses $m_u=m_d=0.33$ GeV, $m_s=0.5$ GeV and
the parameters of the linear potential $A=0.18$ GeV$^2$ and $B=-0.3$ GeV
have the usual values of quark models.  The value of the mixing
coefficient of vector and scalar confining potentials $\varepsilon=-1$
has been determined from the consideration of charmonium radiative
decays \cite{efg}. 
Finally, the universal Pauli interaction constant $\kappa=-1$ has been
fixed from the analysis of the fine splitting of heavy quarkonia ${
}^3P_J$- states \cite{efg}. In this case, the long-range chromomagnetic
interaction of quarks, which is proportional to $(1+\kappa)$, vanishes
in accordance with the flux-tube model.

\section{Quasipotential of the light quark-antiquark interaction}
\label{sec:qpot}

The quasipotential (\ref{qpot}) can  be used for arbitrary quark
masses.  The substitution 
of the Dirac spinors (\ref{spinor}) into (\ref{qpot}) results in an extremely
nonlocal potential in the configuration space. Clearly, it is very hard to 
deal with such potentials without any additional transformations.
 In order to simplify the relativistic $q\bar q$ potential, we make the
following replacement in the Dirac spinors:
\begin{equation}
  \label{eq:sub}
  \epsilon_{1,2}(p)=\sqrt{m_{1,2}^2+{\bf p}^2} \to E_{1,2}
\end{equation}
(see the discussion of this point in \cite{egf,lmes}).  This substitution
makes the Fourier transformation of the potential (\ref{qpot}) local.
 Calculating the potential, we keep only  operators quadratic
in the momentum acting on $V_{\rm Coul}$, $V^{V,S}_{\rm conf}$  and
replace ${\bf p}^2\to E_{1,2}^2-m_{1,2}^2$ in higher order operators
in accord  with Eq.~(\ref{eq:sub}) preserving the symmetry under the
$(1\leftrightarrow 2)$ exchange.  It is necessary to point out that
such substitutions lead to the quark-antiquark potential which commutes
with operators of the total angular momentum  and the orbital angular momentum. Therefore $J$
and $L$ are good quantum numbers, as in the nonrelativistic
approach. However the nonlinear dependence of the interaction
potential on the meson mass effectively takes into account the
relativistic character of the light quark interaction. Note that the
global features of highly excited light mesons can be well understood
in terms of the relativistic relations involving $J$ as well as 
nonrelativistic relations involving $L$ \cite{afonin,fork}.   

The substitution (\ref{eq:sub})
works well for the confining part of the potential. However, it leads to 
fictitious singularities $1/r^3$ and $\delta^3({\bf r})$  at the origin arising from the  
one-gluon exchange part ($\Delta V_{\rm
  Coul}(r)$), which is absent in the initial potential.
Note that these singularities are not important if they are treated
perturbatively. Since we are not using the  expansion in $v/c$ and
are solving the quasipotential equation with the 
complete relativistic potential, an additional analysis is
required. Such singular contributions emerge, e.g., from the following  terms  
\begin{eqnarray}
  \label{eq:st}
 && \frac{{\bf k}^2}{[\epsilon_i(q)(\epsilon_i(q)+m_i)
\epsilon_i(p)(\epsilon_i(p)+m_i)]^{1/2}}V_{\rm Coul}({\bf k}^2) ,\cr
&&\frac{{\bf k}^2}{[\epsilon_1(q)\epsilon_1(p)
\epsilon_2(q)\epsilon_2(p)]^{1/2}}V_{\rm Coul}({\bf k}^2),
\end{eqnarray}
if we simply replace $\epsilon_{1,2}\to E_{1,2}$. However, the Fourier
transforms of expressions (\ref{eq:st}) are less singular at $r\to
0$. To avoid such fictitious singularities we note that if the binding effects 
are taken into account, it is necessary to replace $\epsilon_{1,2}
\to E_{1,2}-\eta_{1,2}V$, where $V$ is the quark interaction potential
and $\eta_{1,2}=m_{2,1}/(m_1+m_2)$. At small
distances  $r\to 0$, the Coulomb singularity in $V$ dominates
and gives the correct asymptotic behaviour. Therefore, we replace
$\epsilon_{1,2} \to E_{1,2}-\eta_{1,2}V_{\rm Coul}$  in  the Fourier
transforms of terms (\ref{eq:st}) (cf. \cite{bs}). We used a
similar regularization of singularities in the analysis of
heavy-light meson spectra \cite{egf}. Finally, we ignore the annihilation
terms in the quark potential since they contribute only in the
isoscalar channels and are suppressed in the $s\bar s$ vector channel.

The resulting $q\bar q$ potential then reads
\begin{equation}
  \label{eq:v}
  V(r)= V_{\rm SI}(r)+ V_{\rm SD}(r),
\end{equation}
where the spin-independent potential has the form 
\begin{eqnarray}
  \label{eq:vsi}
  V_{\rm SI}(r)&=&V_{\rm Coul}(r)+V_{\rm conf}(r)+
\frac{(E_1^2-m_1^2+E_2^2-m_2^2)^2}{4(E_1+m_1)(E_2+m_2)}\Biggl\{
\frac1{E_1E_2}V_{\rm Coul}(r)\cr
&& +\frac1{m_1m_2}\Biggl(1+(1+\kappa)\Biggl[(1+\kappa)\frac{(E_1+m_1)(E_2+m_2)}
{E_1E_2}\cr
&&-\left(\frac{E_1+m_1}{E_1}+\frac{E_1+m_2}{E_2}\right)\Biggr]\Biggr)
V^V_{\rm conf}(r)
+\frac1{m_1m_2}V^S_{\rm conf}(r)\Biggr\}\cr
&&+\frac14\left(\frac1{E_1(E_1+m_1)}\Delta
\tilde V^{(1)}_{\rm Coul}(r)+\frac1{E_2(E_2+m_2)}\Delta
\tilde V^{(2)}_{\rm Coul}(r)\right)\cr
&&-\frac14\left[\frac1{m_1(E_1+m_1)}+\frac1{m_2(E_2+m_2)}-(1+\kappa)
\left(\frac1{E_1m_1}+\frac1{E_2m_2}\right)\right]\Delta V^V_{\rm
conf}(r)\cr
&&+\frac{(E_1^2-m_1^2+E_2^2-m_2^2)}{8m_1m_2(E_1+m_1)(E_2+m_2)} 
\Delta V^S_{\rm conf}(r)+\frac1{E_1E_2}\frac{{\bf L}^2}{2r}\bar V_{\rm Coul}'(r), 
\end{eqnarray}
and the spin-dependent potential is given by
\begin{equation}
  \label{eq:vsd}
   V_{\rm SD}(r)=a_1\, {\bf L}{\bf S}_1+a_2\, {\bf L}{\bf S}_2+
b \left[-{\bf S}_1{\bf S}_2+\frac3{r^2}({\bf S}_1{\bf r})({\bf
    S}_2{\bf r})\right]+ c\, {\bf S}_1{\bf S}_2+ d\, ({\bf L}{\bf
  S}_1) ({\bf L}{\bf S}_2),
\end{equation}
\begin{eqnarray}
\label{eq:a1}
a_1&=&\frac1{2E_1E_2}\Biggl\{\left(2+\frac{2m_2}{E_1+m_1}\right)\frac1{r}
  \bar V_{\rm Coul}'(r)-\frac{2E_2}{E_1+m_1}\frac1{r} V_{\rm conf}'(r)-\left(1+\frac{2m_2}{E_1+m_1}\right)\cr
&&\times\left(\frac{E_1-m_1}{2m_1}
-(1+\kappa)\frac{E_1+m_1}{2m_1}\right)\frac2{r}V'^V_{\rm
    conf}(r) 
+\left(\frac{E_1-m_1}{E_2+m_2}+ \frac{E_2-m_2}{E_1+m_1}\right)\frac1{r}V'^V_{\rm
    conf}(r)\Biggr\}\cr
&&+\frac1{4E_1E_2(E_1+m_1)(E_2+m_2)} \Biggl[\frac1{r}\hat V_{\rm
  Coul}'''(r)+\frac1{r}V'''^S_{\rm conf}(r)+ \left(\frac{E_1}{m_1}
-2(1+\kappa)\frac{E_1+m_1}{2m_1}\right)\cr
&&\times\left(\frac{E_2}{m_2}-2(1+\kappa)\frac{E_2+m_2}{2m_2}\right)
\frac1{r}V'''^V_{\rm  conf}(r)\Biggr],\\
a_2&=&a_1 (1\leftrightarrow 2), \label{eq:a2}\\
\label{eq:b}
b&=&\frac1{3E_1E_2}\Biggl[\frac1{r} \bar V_{\rm Coul}'(r) -\bar V_{\rm
  Coul}''(r) +\left(\frac{E_1-m_1}{2m_1}-(1+\kappa)\frac{E_1+m_1}{2m_1}\right)\cr 
&&\times
\left(\frac{E_2-m_2}{2m_2}-(1+\kappa)\frac{E_2+m_2}{2m_2}\right)
\left(\frac1{r}V'^V_{\rm  conf}(r)-V''^V_{\rm  conf}(r)\right)\Biggr],\\
  \label{eq:c}
  c&=&\frac2{3E_1E_2}\Biggl[\Delta \bar V_{\rm Coul}(r)
+\left(\frac{E_1-m_1}{2m_1}-(1+\kappa)\frac{E_1+m_1}{2m_1}\right)\cr
&&\times
\left(\frac{E_2-m_2}{2m_2}-(1+\kappa)\frac{E_2+m_2}{2m_2}\right)
\Delta V^V_{\rm conf}(r)\Biggr]\\
\label{eq:d}
d&=&-\frac1{E_1E_2(E_1+m_1)(E_2+m_2)}\frac1{r^2}\left[\hat V_{\rm
    Coul}'(r) -\hat V_{\rm Coul}''(r)+\frac1{r}\hat V'_{\rm
    conf}(r)-\hat V''_{\rm  conf}(r)\right],
\end{eqnarray}

\noindent with
\begin{eqnarray}
  \label{eq:tv}
V_{\rm Coul}(r)&=&-\frac43\frac{\alpha_s}{r},\cr
\tilde V^{(i)}_{\rm Coul}(r)&=&V_{\rm Coul}(r)\frac1{\displaystyle\left(1+
\eta_i\frac43\frac{\alpha_s}{E_i}\frac1{r}\right)\left(1+
\eta_i\frac43\frac{\alpha_s}{E_i+m_i}\frac1{r}\right)},\qquad (i=1,2),\cr
  \bar V_{\rm Coul}(r)&=&V_{\rm Coul}(r)\frac1{\displaystyle\left(1+
\eta_1\frac43\frac{\alpha_s}{E_1}\frac1{r}\right)\left(1+
\eta_2\frac43\frac{\alpha_s}{E_2}\frac1{r}\right)}, 
\qquad \eta_{1,2}=\frac{m_{2,1}}{m_1+m_2},\qquad\cr
 \hat V(r)&=&\frac{V(r)}{\displaystyle\left(1+
\eta_1\frac43\frac{\alpha_s}{E_1}\frac1{r}\right)\left(1+
\eta_1\frac43\frac{\alpha_s}{E_1+m_1}\frac1{r}\right)\left(1+
\eta_2\frac43\frac{\alpha_s}{E_2}\frac1{r}\right)\left(1+
\eta_2\frac43\frac{\alpha_s}{E_2+m_2}\frac1{r}\right)}.\qquad
\end{eqnarray}
Here we put  $\alpha_s\equiv\alpha_s(\mu_{12}^2)$ with $\mu_{12}=2m_1
m_2/(m_1+m_2)$. We adopt for $\alpha_s(\mu^2)$ the
simplest model with freezing \cite{bvb}, namely
\begin{equation}
  \label{eq:alpha}
  \alpha_s(\mu^2)=\frac{4\pi}{\displaystyle\beta_0
\ln\frac{\mu^2+M_B^2}{\Lambda^2}}, \qquad \beta_0=11-\frac23n_f,
\end{equation}
where the background mass is $M_B=2.24\sqrt{A}=0.95$~GeV \cite{bvb}, and
$\Lambda=413$~MeV was fixed from fitting the $\rho$
mass.~\footnote{The definition (\ref{eq:alpha}) of $\alpha_s$ can be
  easily matched with the $\alpha_s$ used for heavy quarkonia
  \cite{efg} at the scale about $m_c$.} We put the
number of flavours $n_f=2$ for $\pi$, 
$\rho$, $K$, $K^*$ and $n_f=3$ for $\phi$. As a result we obtain
$\alpha_s(\mu_{ud}^2)=0.730$, $\alpha_s(\mu_{us}^2)=0.711$ and
$\alpha_s(\mu_{ss}^2)=0.731$.  Note that the other popular
parametrisation of $\alpha_s$ with freezing \cite{shirkov} gives close
values.   

\section{Results and discussion}
\label{sec:rd}

The calculated masses of light unflavoured and strange mesons are
given in Tables~\ref{tab:nsmm} and \ref{tab:smm}. They  are confronted
with available experimental data from PDG 
Particle Listings including data from the ``Further States'' Section
\cite{pdg}. We find good agreement of our predictions with
data. Most of the well-established state masses are reproduced in our
model. 

We do not consider the mixing of states in the isoscalar sector. Therefore the
predictions in Table~\ref{tab:nsmm} are given for the pure $q\bar q$
and $s\bar s$ states. Such mixing is mostly important in the pseudoscalar
sector. 
We follow the $\eta-\eta'$ mixing scheme proposed in Ref.~\cite{fks}
and take the phenomenological values of the mixing angle $\phi=38^\circ$ and the
decay constant ratio $y\equiv f_q/f_s=0.81$. Using our values for the
mass of  $M_{\eta_{s\bar s}}$ and the pion mass  we get
$M_\eta=573$~MeV and $M_{\eta'}=989$~MeV close to the  measured masses
$M_{\eta}^{\rm exp}=547.853\pm0.0024$~MeV and $M_{\eta'}^{\rm
  exp}=957.66\pm0.24$~MeV \cite{pdg}.
The experiment shows that the vector and excited isoscalar
light mesons are almost ideally mixed and therefore can be roughly
considered as  pure  $q\bar q$ and $s\bar s$ states. Indeed we find
reasonable agreement of our prediction with experiment in the
isoscalar sector.

\begin{longtable}{@{ \ }c@{ \ }c@{ \ \ }c@{ \ }c@{ \ }c@{ \ }c@{ \ }c@{ \ \
    }c@{ \ }c@{ \ }c@{ \ }}
\caption{Masses of excited light ($q=u,d$) unflavored mesons
   (in MeV).} 
   \label{tab:nsmm}\\
\hline
\hline\vspace*{-0.5cm}
\\[1pt]
&&Theory
&\multicolumn{4}{l}{\underline{\hspace{2.1cm}Experiment\hspace{2.1cm}}}&
Theory&\multicolumn{2}{l}{\underline{\hspace{.6cm}Experiment\hspace{.6cm}}}\\
$n^{2S+1}L_J$&$J^{PC}$&$q\bar q$ & $I=1$&mass& $I=0$&mass&$s\bar
s$ & $I=0$&mass\\[2pt]
\hline
\endfirsthead
\caption[]{(continued)}\\
\hline\hline
&&Theory
&\multicolumn{4}{l}{\underline{\hspace{2.1cm}Experiment\hspace{2.1cm}}}&
Theory&\multicolumn{2}{l}{\underline{\hspace{.6cm}Experiment\hspace{.6cm}}}\\
$n^{2S+1}L_J$&$J^{PC}$&$q\bar q$ & $I=1$&mass& $I=0$&mass&$s\bar
s$ & $I=0$&mass\\[2pt]
\hline
\endhead
\hline
\hline
\endfoot
\endlastfoot
$1^1S_0$& $0^{-+}$&154& $\pi$&139.57&& &743&&\\
$1^3S_1$& $1^{--}$&776& $\rho$&775.49(34)&$\omega$&782.65(12)
&1038&$\varphi$& 1019.455(20)\\
$1^3P_0$& $0^{++}$&1176& $a_0$&1474(19)&$f_0$&1200-1500 &1420&$f_0$&1505(6)\\
$1^3P_1$& $1^{++}$&1254& $a_1$&1230(40)&$f_1$&1281.8(6)
&1464&$f_1$&1426.4(9)\\
$1^3P_2$& $2^{++}$&1317& $a_2$&1318.3(6)&$f_2$&1275.1(12)
&1529&$f_2'$&1525(5)\\
$1^1P_1$& $1^{+-}$&1258& $b_1$&1229.5(32)&$h_1$&1170(20)
&1485&$h_1$&1386(19)\\
$2^1S_0$& $0^{-+}$&1292& $\pi$&1300(100)&$\eta$&1294(4) &1536&$\eta$&1476(4)\\
$2^3S_1$& $1^{--}$&1486& $\rho$&1465(25)&$\omega$&1400-1450
&1698&$\varphi$& 1680(20)\\
$1^3D_1$& $1^{--}$&1557& $\rho$&1570(70)&$\omega$ &1670(30) &1845&&\\
$1^3D_2$& $2^{--}$&1661& &&& &1908&&\\
$1^3D_3$& $3^{--}$&1714& $\rho_3$&1688.8(21)&$\omega_3$&1667(4)
&1950&$\varphi_3$&1854(7)\\
$1^1D_2$& $2^{-+}$&1643& $\pi_2$&1672.4(32)&$\eta_2$&1617(5)
&1909&$\eta_2$&1842(8)\\
$2^3P_0$& $0^{++}$&1679& &&$f_0$&1724(7) &1969&&\\
$2^3P_1$& $1^{++}$&1742& $a_1$&1647(22)&&
&2016&$f_1$&1971(15)\\
$2^3P_2$& $2^{++}$&1779& $a_2$&1732(16)&$f_2$&1755(10)
&2030&$f_2$&2010(70)\\
$2^1P_1$& $1^{+-}$&1721& &&&
&2024&&\\
$3^1S_0$& $0^{-+}$&1788& $\pi$&1816(14)&$\eta$&1756(9) &2085&$\eta$&2103(50)\\
$3^3S_1$& $1^{--}$&1921& $\rho$&1909(31)&$\omega$&1960(25)
&2119&$\varphi$&2175(15) \\
$1^3F_2$& $2^{++}$&1797& &&$f_2$&1815(12) &2143&$f_2$&2156(11)\\
$1^3F_3$& $3^{++}$&1910&$a_3$ &1874(105)&&
&2215&$f_3$& 2334(25)\\
$1^3F_4$& $4^{++}$&2018& $a_4$&2001(10)&$f_4$&2018(11)
&2286&&\\
$1^1F_3$& $3^{+-}$&1884& &&&
&2209&$h_3$&2275(25)\\
$2^3D_1$& $1^{--}$&1895& $\rho$&1909(31)&& &2258&$\omega$&2290(20)\\
$2^3D_2$& $2^{--}$&1983&$\rho_2$ &1940(40)&$\omega_2$&1975(20) &2323&&\\
$2^3D_3$& $3^{--}$&2066& &&&
&2338&&\\
$2^1D_2$& $2^{-+}$&1960& $\pi_2$&1974(84)&$\eta_2$&2030(20)
&2321&&\\
$3^3P_0$& $0^{++}$&1993&$a_0$ &2025(30)&$f_0$&1992(16) &2364&$f_0$&2314(25)\\
$3^3P_1$& $1^{++}$&2039&$a_1$ &2096(123)&&
&2403&&\\
$3^3P_2$& $2^{++}$&2048&$a_2$ &2050(42)&$f_2$&2001(10)
&2412&$f_2$&2339(60)\\
$3^1P_1$& $1^{+-}$&2007&$b_1$ &1960(35)&$h_1$&1965(45)
&2398&&\\
$4^1S_0$& $0^{-+}$&2073& $\pi$&2070(35)&$\eta$&2010(50) &2439&&\\
$4^3S_1$& $1^{--}$&2195& $\rho$&2265(40)&$\omega$&2205(30)
&2472&& \\
$1^3G_3$& $3^{--}$&2002& $\rho_3$&1982(14)&$\omega_3$ &1945(20) &2403&&\\
$1^3G_4$& $4^{--}$&2122&$\rho_4$ &2230(25)&$\omega_4$&2250(30) &2481&&\\
$1^3G_5$& $5^{--}$&2264& $\rho_5$&2300(45)&$\omega_5$&2250(70)
&2559&&\\
$1^1G_4$& $4^{-+}$&2092& &&&
&2469&&\\
$3^3D_1$& $1^{--}$&2168& $\rho$&2149(17)&& &2607&\\
$3^3D_2$& $2^{--}$&2241&$\rho_2$ &2225(35)&$\omega_2$&2195(30) &2667&&\\
$3^3D_3$& $3^{--}$&2309&$\rho_3$ &2300(60)&$\omega_3$&2278(28)
&2727&\\
$3^1D_2$& $2^{-+}$&2216& $\pi_2$&2245(60)&$\eta_2$&2248(20)
&2662&&\\ \pagebreak
$2^3F_2$& $2^{++}$&2091&$a_2$ &2100(20)&$f_2$&2141(12)
&2514&&\\
$2^3F_3$& $3^{++}$&2191&$a_3$ &2070(20)&&
&2585&&\\
$2^3F_4$& $4^{++}$&2284& &&$f_4$&2320(60)
&2657&&\\
$2^1F_3$& $3^{+-}$&2164&$b_3$ &2245(50)&&
&2577&&\\
$4^3P_0$& $0^{++}$&2250& &&$f_0$&2189(13)
&2699&&\\
$4^3P_1$& $1^{++}$&2286& $a_1$&2270(50)&$f_1$&2310(60) &2729&&\\
$4^3P_2$& $2^{++}$&2297&$a_2$&2280(30) &$f_2$&2297(28) &2734&&\\
$4^1P_1$& $1^{+-}$&2264&$b_1$&2240(35)&$h_1$&2215(40) &2717&&\\
$2^3G_3$& $3^{--}$&2267&$\rho_3$ &2260(20)&$\omega_3$&2255(15)
&2743&&\\
$2^3G_4$& $4^{--}$&2375& &&&
&2819&&\\
$2^3G_5$& $5^{--}$&2472& &&&
&2894&&\\
$2^1G_4$& $4^{-+}$&2344&$\pi_4$ &2250(15)&$\eta_4$&2328(30)
&2806&&\\
$5^1S_0$& $0^{-+}$&2385& $\pi$&2360(25)&$\eta$&2320(15) &2749&&\\
$5^3S_1$& $1^{--}$&2491& &&
&&2782&& \\
$1^3H_4$& $4^{++}$&2234&$a_4$ &2237(5)&$f_J$& 2231.1(35)
&2634&&\\
$1^3H_5$& $5^{++}$&2359& &&&
&2720&&\\
$1^3H_6$& $6^{++}$&2475&$a_6$ &2450(130)&$f_6$&2465(50)
&2809&&\\
$1^1H_5$& $5^{+-}$&2328&&&&
&2706&& \\[2pt]
\hline
\hline
\end{longtable}

The strange meson states ($L_L$) with $J=L$, given in Table~ \ref{tab:smm},
are the mixtures of spin-triplet ($^3L_L$)  and spin-singlet ($^1L_L$)
states:
\begin{eqnarray}
  \label{eq:mix}
  K_J&=&K(^1L_L)\cos\varphi+K(^3L_L)\sin\varphi, \cr
 K'_J&=&-K(^1L_L)\sin\varphi+K(^3L_L)\cos\varphi, \qquad J=L=1,2,3\dots
\end{eqnarray}

\noindent where $\varphi$ is a mixing angle.
  Such mixing occurs due to the nondiagonal spin-orbit and
tensor terms in Eq.~(\ref{eq:vsd}). The masses of  physical states were obtained
by diagonalising the mixing terms. The found values of mixing angle
$\varphi$ are the 
following: $1P$ $43.8^\circ$, $2P$ $44.6^\circ$,  $3P$ $44.8^\circ$,
$1D$ $44.2^\circ$, $2D$ $44.5^\circ$, $1F$ $44.3^\circ$, $1G$
$44.3^\circ$. These values show that physical $K_J$ mesons are
nearly equal mixtures of the corresponding spin-singlet $K(^1L_L)$ and
spin-triplet $K(^3L_L)$ states which is in good agreement with the experimental
data \cite{pdg}. 

\begin{table}
  \caption{Masses of excited strange mesons (in MeV).} 
  \label{tab:smm}
\begin{ruledtabular}
\begin{tabular}{cccccccccc}
&&Theory
&\multicolumn{2}{l}{\underline{\hspace{0.7cm}Experiment\hspace{0.7cm}}}&
& &
Theory&\multicolumn{2}{l}{\underline{\hspace{.7cm}Experiment\hspace{.7cm}}}\\
$n^{2S+1}L_J$&$J^{P}$& $q\bar s$& $I=1/2$&mass&$n^{2S+1}L_j$&$J^{P}$&$q\bar s$ &$I=1/2$&mass\\
\hline
$1^1S_0$& $0^{-}$&482& $K$&493.677(16)&$3^1S_0$& $0^{-}$&2065& &\\
$1^3S_1$& $1^{-}$&897& $K^*$&891.66(26)&$3^3S_1$& $1^{-}$&2156& &\\
$1^3P_0$& $0^{+}$&1362& $K_0$&1425(50)&$2^3D_1$&  $1^{-}$&2063&&\\
$1^3P_2$& $2^{+}$&1424& $K_2^*$&1425.6(15)&$2^3D_3$& $3^{-}$
&2182&&\\
$1P_1$& $1^{+}$&1412& $K_1$&1403(7)&$2D_2$& $2^{-}$
&2163&$K_2$&2247(17)\\
$1P_1$& $1^{+}$&1294& $K_1$&1272(7)&$2D_2$& $2^{-}$
&2066&&\\
$2^1S_0$& $0^{-}$&1538&&&$3^3P_0$& $0^{+}$&2160&& \\
$2^3S_1$& $1^{-}$&1675& $K^*$&&$3^3P_2$& $2^{+}$
&2206&&\\
$1^3D_1$& $1^{-}$&1699& $K^*$&1717(27)&$3P_1$& $1^{+}$&2200&&\\
$1^3D_3$& $3^{-}$&1789& $K^*_3$&1776(7)&$3P_1$& $1^{+}$
&2164&&\\
$1D_2$& $2^{-}$&1824&$K_2$ &1816(13)&$1^3G_3$& $3^{-}$ &2207&&\\
$1D_2$& $2^{-}$&1709& $K_2$&1773(8)&$1^3G_5$& $5^{-}$
&2356&$K_5^*$&2382(24)\\
$2^3P_0$& $0^{+}$&1791& &&$1G_4$& $4^{-}$
&2285&&\\
$2^3P_2$& $2^{+}$&1896&&&$1G_4$& $4^{-}$
&2255&&\\
$2P_1$& $1^{+}$&1893& & &$2^3F_4$& $4^{+}$ &2436&&\\
$2P_1$& $1^{+}$&1757&$K_1$ &1650(50)&$2F_3$& $3^{+}$
&2348&$K_3$&2324(24)\\
$1^3F_2$& $2^{+}$&1964& $K_2^*$&1973(26)&$2^3G_5$& $5^{-}$
&2656&&\\
$1^3F_4$& $4^{+}$&2096& $K^*_4$&2045(9)&$2G_4$& $4^{-}$
&2575&$K_4$&2490(20)\\
$1F_3$& $3^{+}$&2080& &&\\
$1F_3$& $3^{+}$&2009& &&\\
\end{tabular}
 \end{ruledtabular}
\end{table}

The scalar sector presents a special interest due to its complexity
and the abundance
of experimentally observed light states. We see from
Table~\ref{tab:nsmm} that the masses of 
the lightest $q\bar q$ scalar mesons  have values about 1200 MeV.  
This confirms the conclusion of our
recent consideration \cite{ltetr} that light scalar mesons,
$f_0(600)$ ($\sigma$), $K^*_0(800)$ ($\kappa$), $f_0(980)$ and $a_0(980)$, with
masses below 1 GeV should be described as light tetraquarks consisting
from scalar diquark and antidiquark. Moreover the predicted masses of the
scalar tetraquarks composed from axial-vector  diquark and antidiquark  \cite{ltetr} 
have masses in the same range as the lowest $q\bar q$ scalar
mesons. Therefore mixing between these states can occur, e.g. due to
the instanton-induced mixing terms \cite{schechter,maiani}. The
obtained results for the masses indicate that $a_0(1450)$ should be
predominantly a tetraquark state which predicted \cite{ltetr} mass 1480 MeV is within
experimental error bars $M_{a_0(1450)}=1474\pm19$~MeV. The 
exotic scalar state $X(1420)$ from the ``Further States'' Section
could be its isotensor partner. On the other 
hand $s\bar q(1^3P_0)$  interpretation is favored for $K_0^*(1430)$
(see Table~\ref{tab:smm}). This picture naturally explains the
experimentally observed proximity of masses of the unflavoured
$a_0(1450)$ and $f_0(1500)$ with the strange $K_0^*(1430)$. Therefore
one could expect an additional isovector predominantly $q\bar
q$ state $a_0$ with the mass about 1200 MeV, though it was not
observed in several experiments.    

It was noted long ago that the light meson Regge trajectories are
almost linear in  $(J,M^2)$ and $(n_r,M^2)$ planes.\\
a) The $(J,M^2)$ Regge trajectory:

\begin{equation}
  \label{eq:reggej}
J=\alpha M^2+\alpha_0;
\end{equation}

\noindent b) The $(n_r,M^2)$ Regge trajectory:

\begin{equation}
  \label{eq:reggen}
n_r\equiv n-1=\beta M^2+\beta_0,
\end{equation}
where $\alpha$, $\beta$ are the slopes and  $\alpha_0$, $\beta_0$ are
intercepts. The relations (\ref{eq:reggej}) and (\ref{eq:reggen})
arise in most models of quark confinement, but with different values
of the slopes. For example, the QCD string with two light quarks
at the ends gives the slopes  \cite{simonov}:
\begin{equation}
  \label{eq:str}
  \alpha=\frac1{2\pi \sigma},\qquad \beta=\frac1{4\pi \sigma},
\end{equation}
where $\sigma$ is the
string tension which is equal to the slope of the
linear confining potential $A$ in Eq.~(\ref{nr}).

On the other hand, the quasiclassical picture for a
light meson, described by the massless Salpeter
equation with a linear confining potential:
\begin{equation}
  \label{eq:slp}
  (2p+Ar)\psi=M\psi,
\end{equation}
gives for the Regge slopes \cite{bicudo} 
\begin{equation}
  \label{eq:sls}
  \alpha=\frac1{8A},\qquad \beta=\frac1{4\pi A},
\end{equation}
implying that
\begin{equation}
  \label{eq:slsr}
  \alpha/\beta=\pi/2.
\end{equation}

\begin{figure}
 \includegraphics[width=13cm]{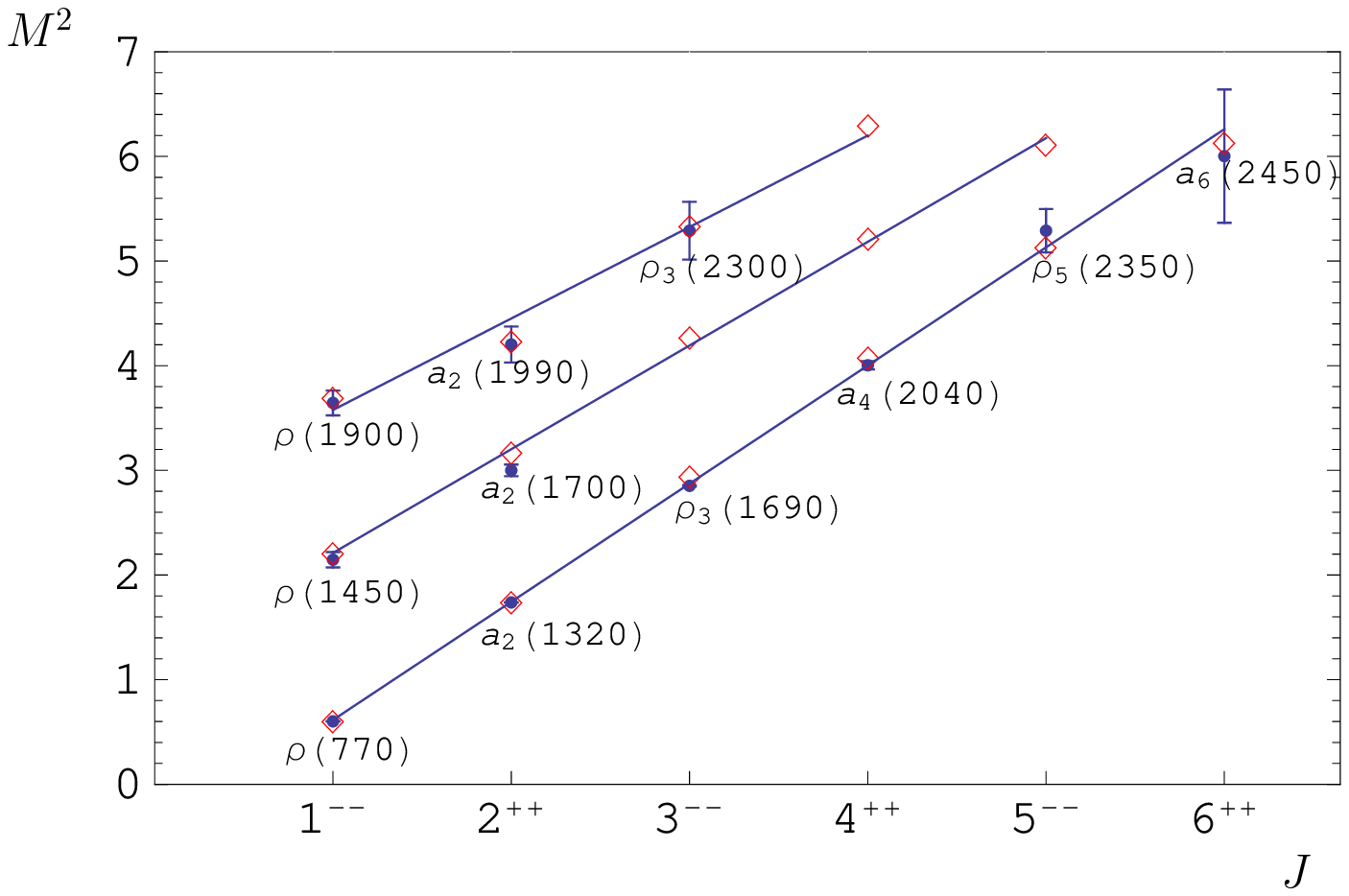}

\caption{\label{fig:rr} Parent and daughter ($J, M^2$) Regge trajectories for
  isovector light mesons  with natural parity ($\rho$). Diamonds are predicted
  masses. Available experimental data are given by dots with error
  bars and particle names. $M^2$ is in GeV$^2$. }
\end{figure}

\begin{figure}
  \includegraphics[width=13cm]{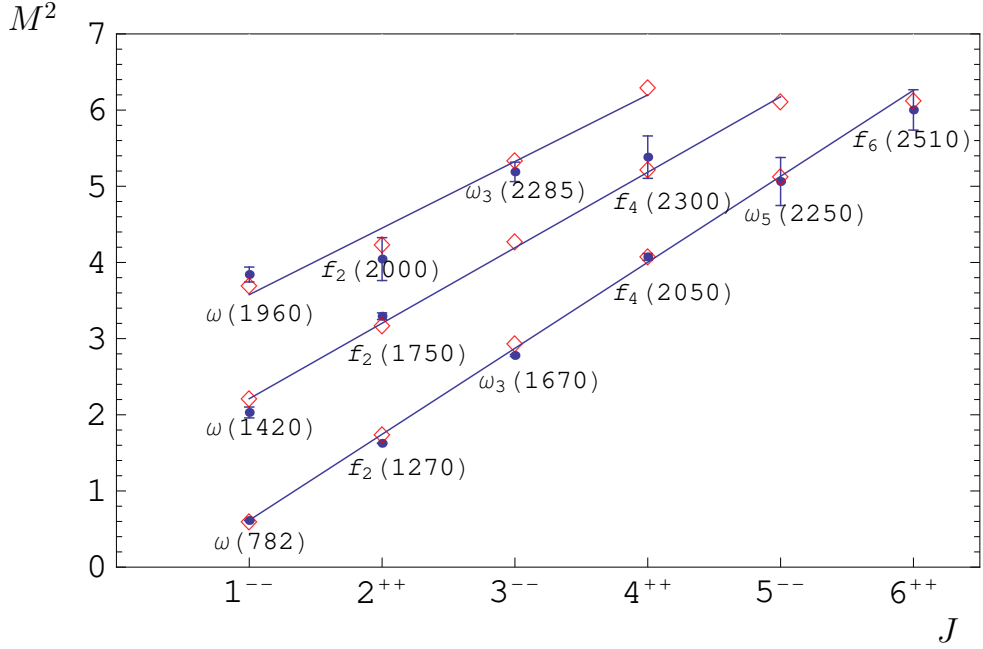}

\caption{\label{fig:0r} Same as in Fig.~\ref{fig:rr} for
  isoscalar light $q\bar q$ mesons with natural parity ($\omega$). }
\end{figure}

\begin{figure}
  \includegraphics[width=13cm]{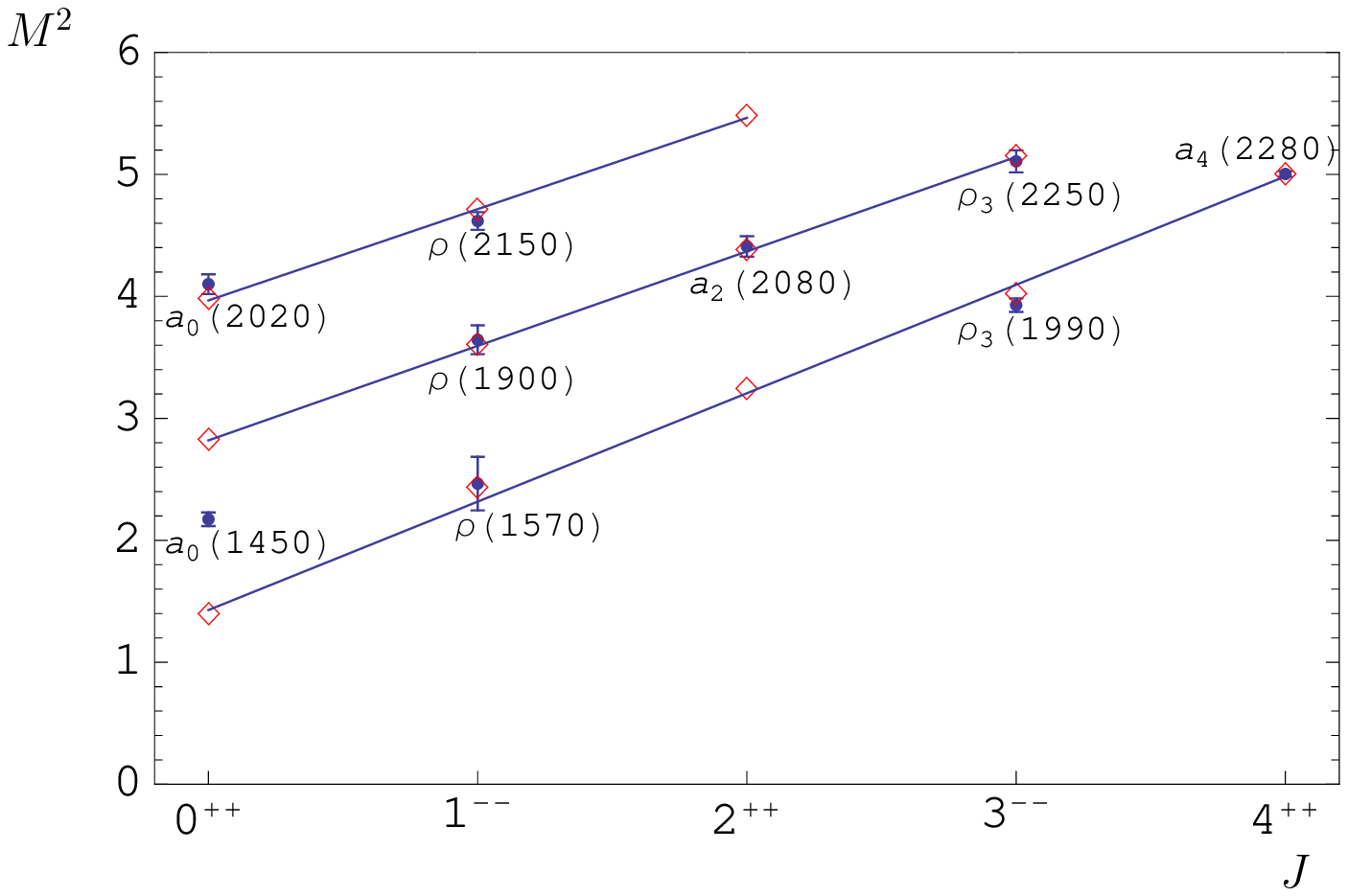}

\caption{\label{fig:a0r} Same as in Fig.~\ref{fig:rr} for
  isovector light $q\bar q$ mesons with natural parity ($a_0$). }
\end{figure}

\begin{figure}
  \includegraphics[width=13cm]{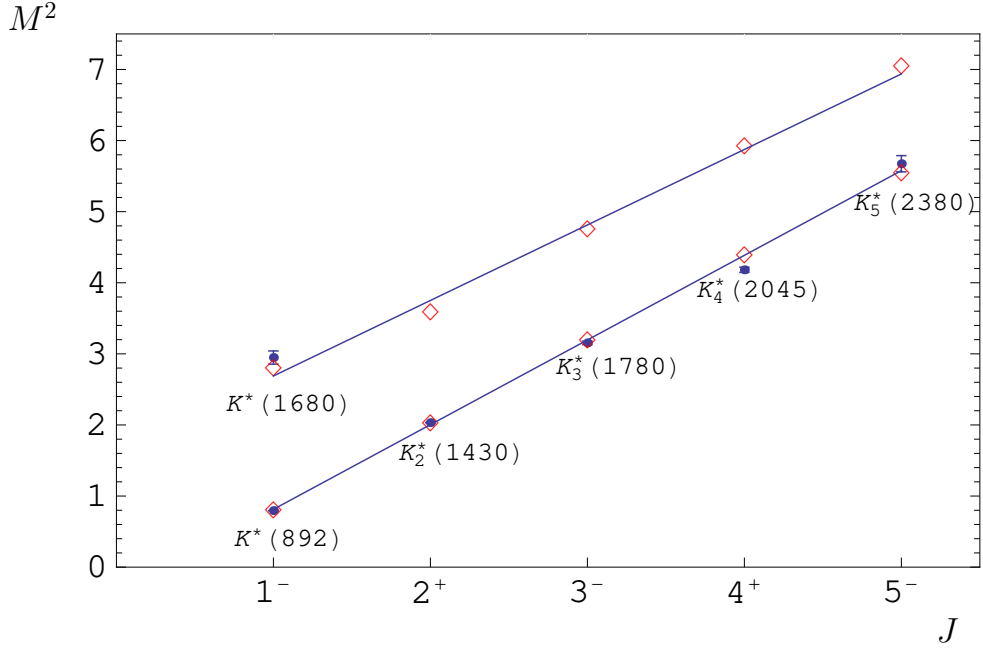}

\caption{\label{fig:ksr} Same as in Fig.~\ref{fig:rr} for
  isodublet light mesons  with natural parity ($K^*$). }
\end{figure}

\begin{figure}
  \includegraphics[width=13cm]{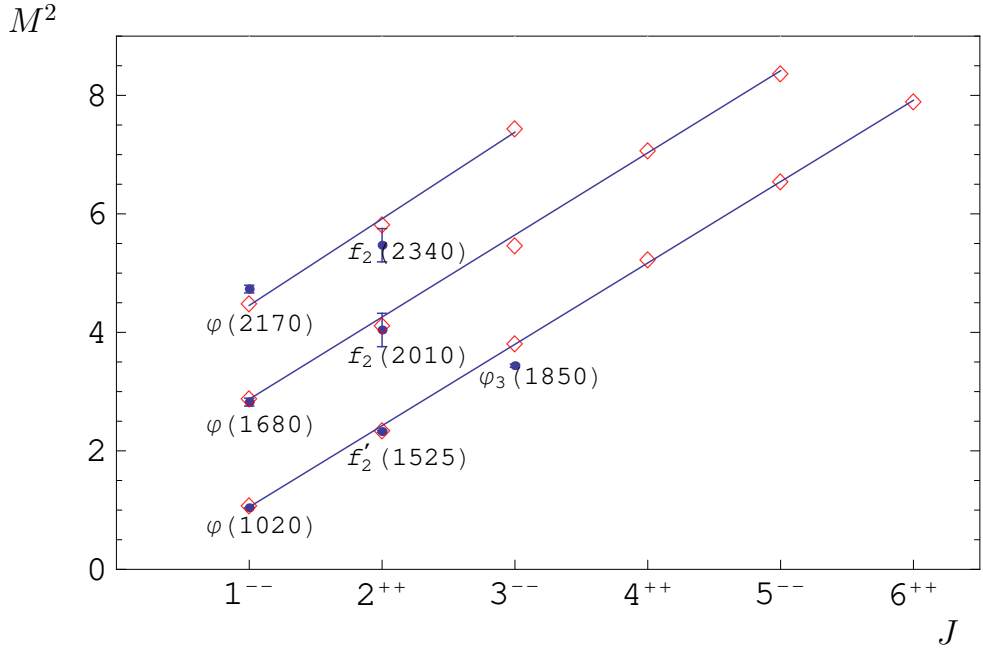}

\caption{\label{fig:phir} Same as in Fig.~\ref{fig:rr} for
  isoscalar light $s\bar s$ mesons  with natural parity ($\varphi$). }
\end{figure}

\begin{figure}
  \includegraphics[width=13cm]{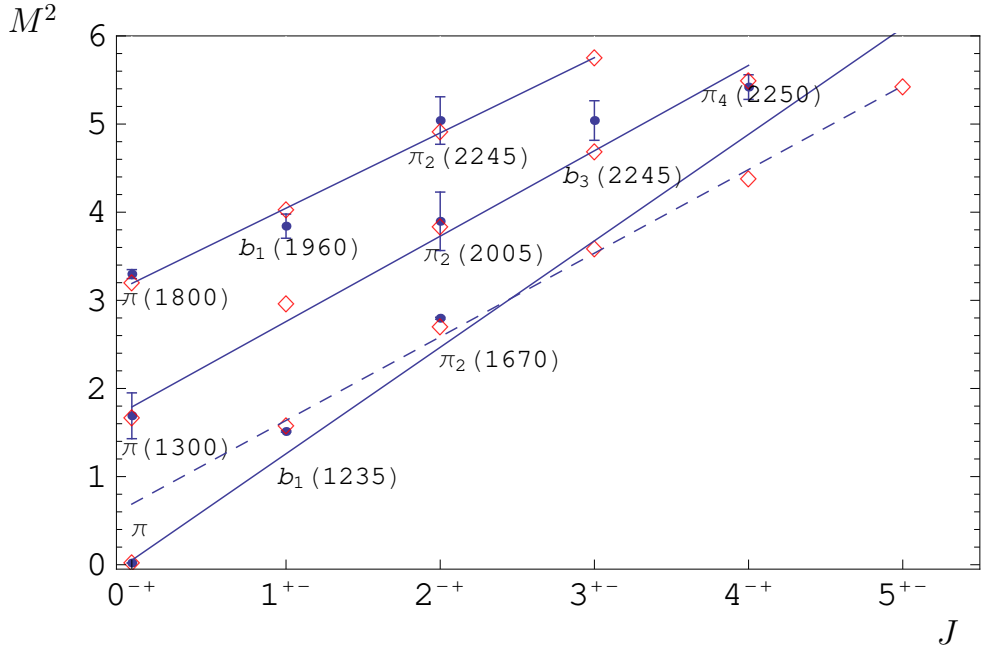}

\caption{\label{fig:pir} Same as in Fig.~\ref{fig:rr} for
  isovector light mesons  with unnatural parity ($\pi$). Dashed line corresponds
to the Regge trajectory, fitted without $\pi$.  }
\end{figure}

\begin{figure}
  \includegraphics[width=13cm]{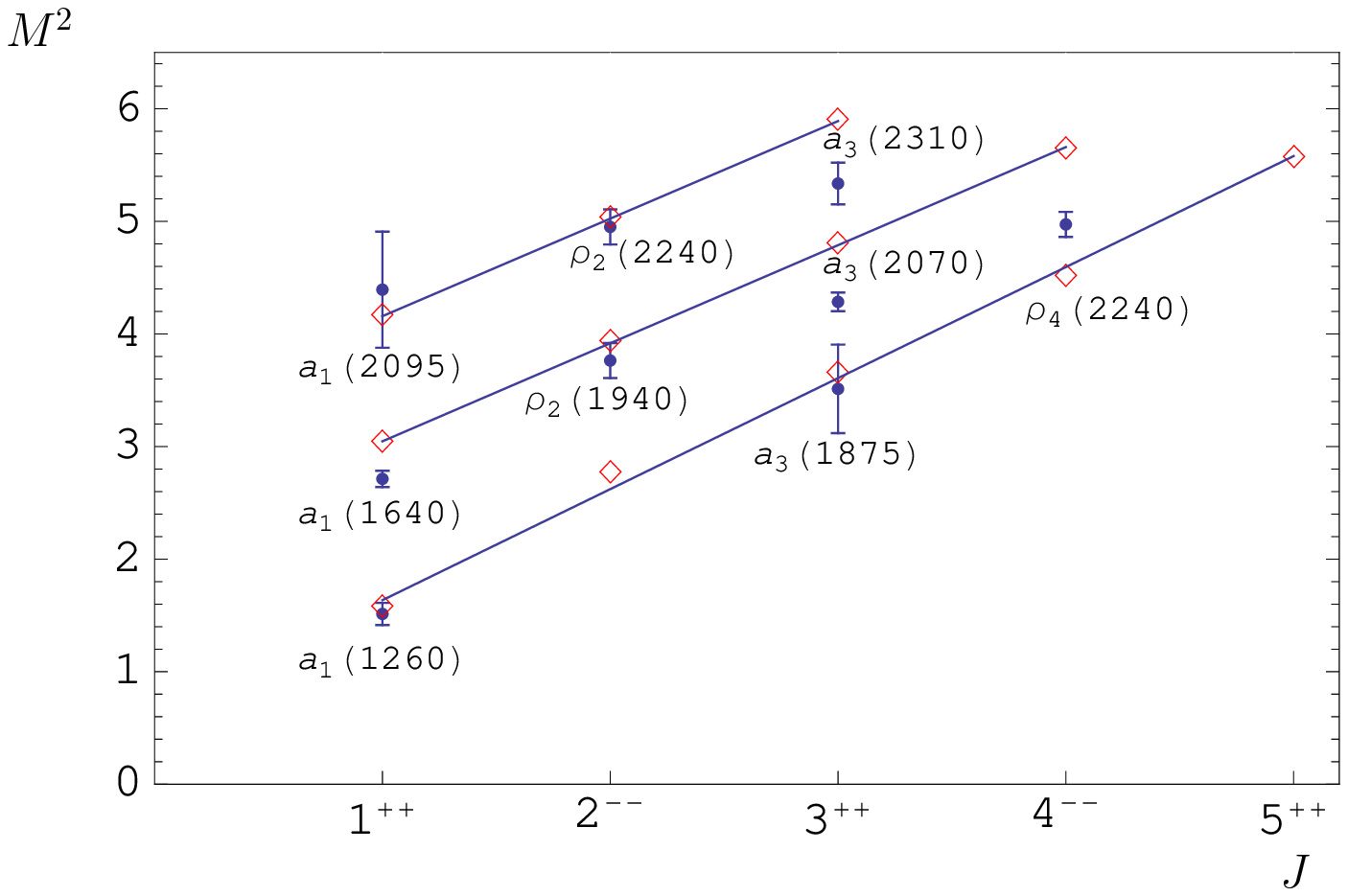}

\caption{\label{fig:a1r} Same as in Fig.~\ref{fig:rr} for
  isovector light $q\bar q$ mesons with unnatural parity ($a_1$). }
\end{figure}

\begin{figure}
\vspace*{-0.2cm}  
\includegraphics[width=13cm]{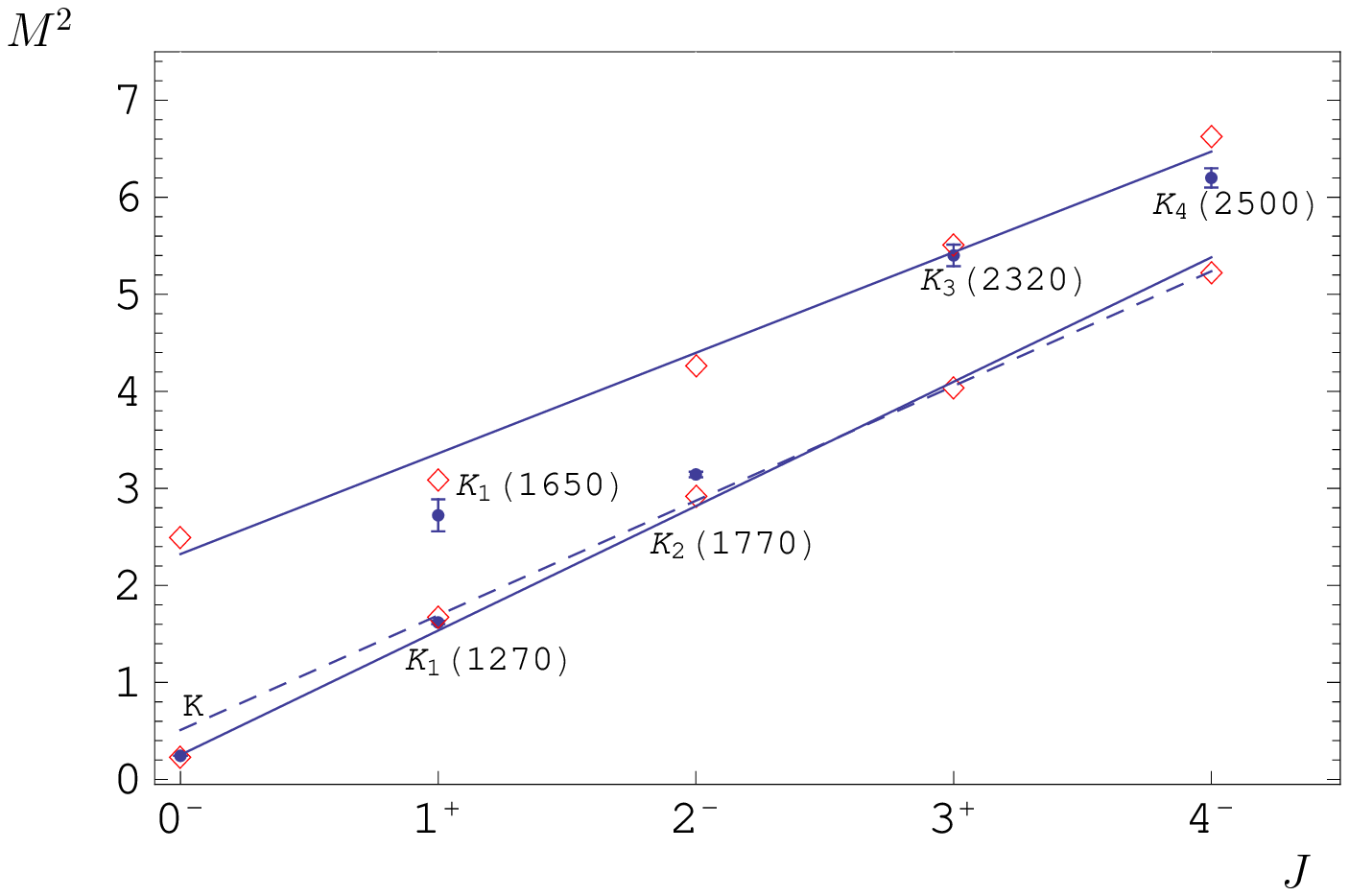}

\caption{\label{fig:kr} Same as in Fig.~\ref{fig:rr} for
  isodublet light mesons  with unnatural parity ($K$). Dashed line corresponds
to the Regge trajectory, fitted without $K$.}
\end{figure}

\begin{figure}
  \includegraphics[width=13cm]{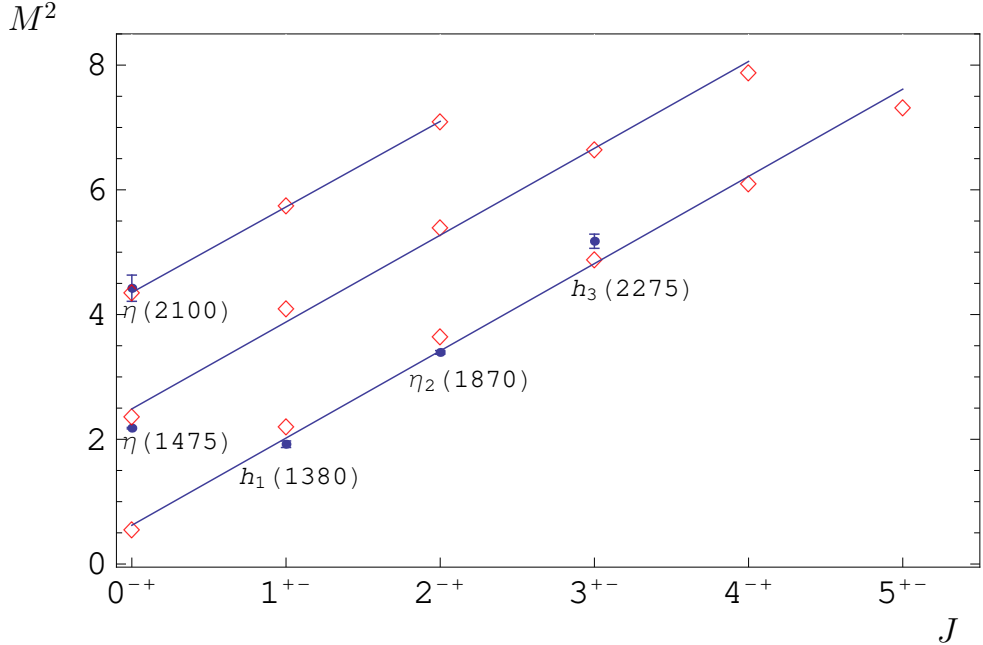}

\caption{\label{fig:phiru} Same as in Fig.~\ref{fig:rr} for
  isoscalar light $s\bar s$ mesons with unnatural parity. The ground
  state with $J=0$ is the mixture of $\eta$ and $\eta'$ with pure $s\bar
  s$ quark content ($\eta_{s\bar s}$). }
\end{figure}

\begin{figure}
  \includegraphics[width=13cm]{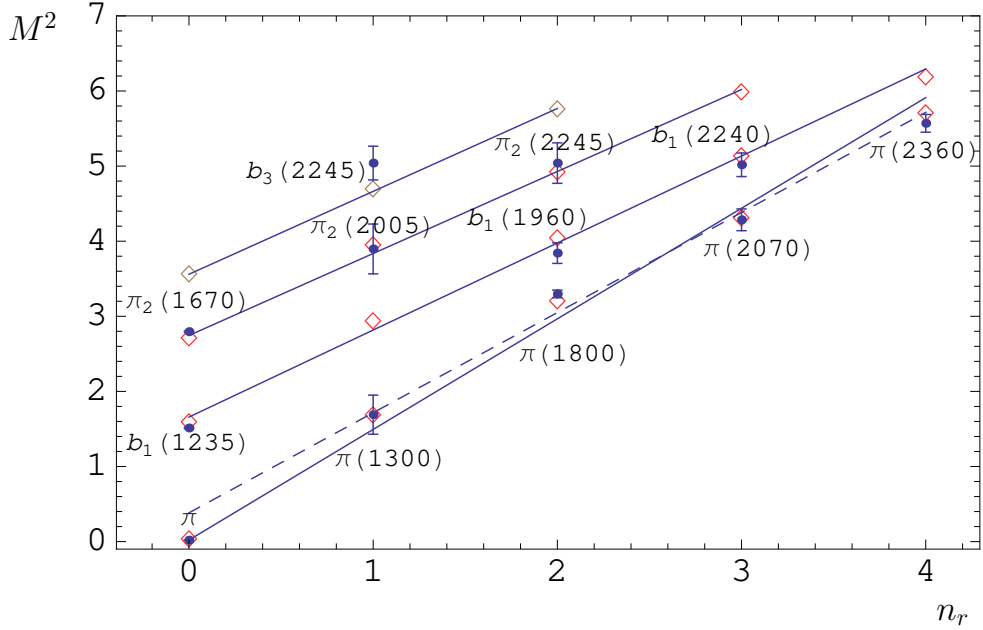}

\caption{\label{fig:pin} The $(n_r,M^2)$ Regge trajectories for
  spin-singlet isovector mesons $\pi$,
  $b_1$, $\pi_2$ and $b_3$  (from bottom to top). Notations are the
  same as in Fig.~\ref{fig:rr}.  The dashed line corresponds 
to the Regge trajectory, fitted without $\pi$.}
\end{figure}

\begin{figure}
  \includegraphics[width=13cm]{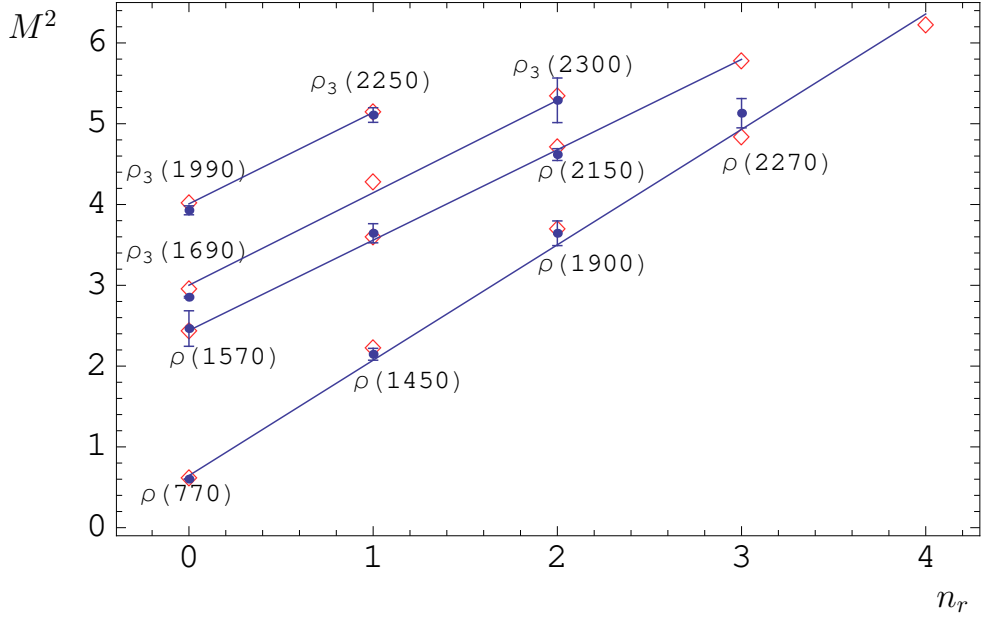}

\caption{\label{fig:rhn} The $(n_r,M^2)$ Regge trajectories for
  spin-triplet isovector mesons 
  $\rho (^3S_1)$, $\rho(^3D_1)$, $\rho_3(^3D_3)$ and  $\rho_3(^3G_3)$  (from bottom to top). Notations are the same as in Fig.~\ref{fig:rr}.  }
\end{figure}

\begin{figure}
  \includegraphics[width=13cm]{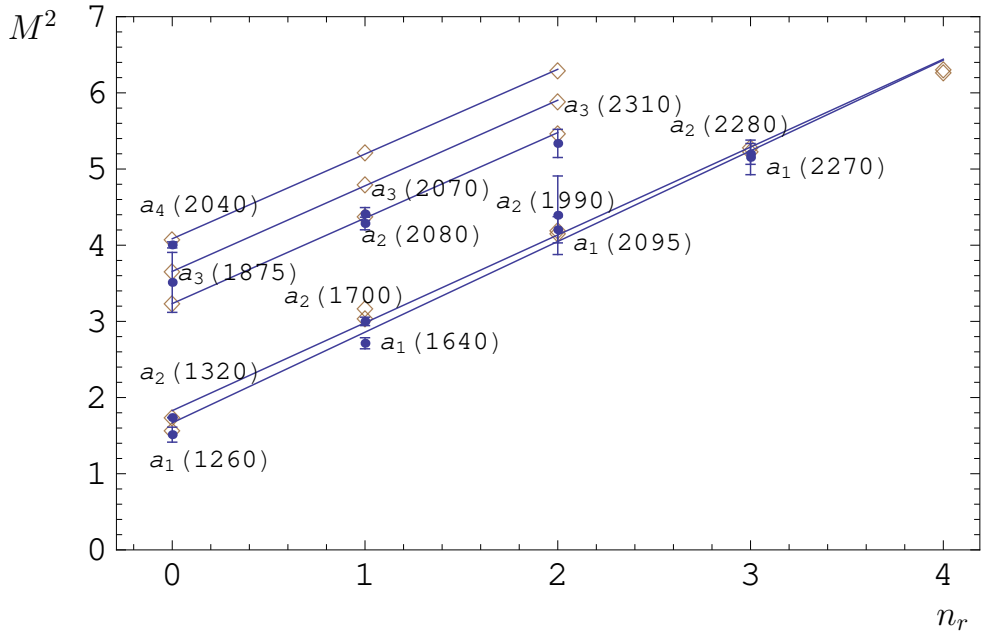}

\caption{\label{fig:an} The $(n_r,M^2)$ Regge trajectories for spin-triplet isovector mesons  $a_1$,
  $a_2(^3P_2)$, $a_2(^3F_2)$, $a_3$ and $a_4(^3F_4)$ (from bottom to top). Notations are the same as in Fig.~\ref{fig:rr}. }
\end{figure}

In Figs.~\ref{fig:rr}-\ref{fig:phir} and
Figs.~\ref{fig:pir}-\ref{fig:phiru} we plot the Regge trajectories in
the ($J, M^2$) plane for mesons with natural ($P=(-1)^J $) and
unnatural ($P=(-1)^{J-1}$) parity,
respectively. The Regge trajectories in the $(n_r,M^2)$ plane are presented
in Figs.~\ref{fig:pin}-\ref{fig:an}. The masses calculated in our
model are shown by diamonds. Available experimental data are given by
dots with error bars and corresponding meson names. 
Straight lines were obtained by a
$\chi^2$ fit of the calculated values. The fitted slopes
and intercepts of the Regge trajectories are given in
Tables~\ref{tab:rtj} and \ref{tab:rtn}. We see that the calculated
light meson masses fit nicely to the linear trajectories in both
planes. These trajectories are almost parallel and equidistant.

\begin{table}
  \caption{Fitted parameters of the $(J,M^2)$ parent and daughter Regge
    trajectories for light mesons 
    with natural and unnatural parity ($q=u,d$).} 
  \label{tab:rtj}
\begin{ruledtabular}
\begin{tabular}{ccccc}
Trajectory& 
\multicolumn{2}{l}{\underline{\hspace{1.6cm}natural
    parity\hspace{1.6cm}}}&
 \multicolumn{2}{l}{\underline{\hspace{1.6cm}unnatural
    parity\hspace{1.6cm}}}\hspace{-.1cm}\\ 
&$\alpha$ (GeV$^{-2}$)& $\alpha_0$&$\alpha$  (GeV$^{-2}$)&$\alpha_0$\\
\hline
$q\bar q$ &$\rho$&&$\pi$\\
parent& $0.887\pm0.008$& $0.456\pm0.018$& $0.828\pm0.057^*$&
$-0.025\pm0.034^*$\\
daughter 1&$1.009\pm0.019$&$-1.232\pm0.074$ & $1.031\pm0.063$&
$-1.846\pm0.217$\\
daughter 2&$1.144\pm0.113$&$-3.092\pm0.540$ & $1.171\pm0.009$&
$-3.737\pm0.042$\\
$q\bar q$ &$a_0$&&$a_1$\\
parent& $1.125\pm0.035$& $-1.607\pm0.104$& $1.014\pm0.036$&
$-0.658\pm0.120$\\
daughter 1&$1.291\pm0.003$&$-3.640\pm0.011$ & $1.148\pm0.012$&
$-2.497\pm0.050$\\
daughter 2&$1.336\pm0.022$&$-5.300\pm0.102$ & $1.154\pm0.014$&
$-3.798\pm0.007$\\
$q\bar s$& $K^*$&&$K$\\
parent& $0.839\pm0.004$&$0.318\pm0.012$ & $0.780\pm0.022^\dag$&
$-0.197\pm0.036^\dag$\\
daughter&$0.942\pm0.046$&$-1.532\pm0.209$&$0.964\pm0.072$&
$-2.240\pm0.296$\\
$s\bar s$&$\varphi$&&$\eta_{s\bar s}$\\
parent&$0.728\pm0.011$&$0.234\pm0.034$&$0.715\pm0.023$
&$-0.444\pm0.068$\\
daughter 1& $0.721\pm0.089$&$-1.072\pm0.047$&$0.718\pm0.032$&
$-1.786\pm0.157$\\ 
 daughter 2& $0.684\pm0.039$&$-2.047\pm0.226$&$0.729\pm0.010$&
$-3.174\pm0.057$\\ 
\end{tabular}
 \end{ruledtabular}
${}^*$ fit without $\pi$: $\alpha=(1.053\pm0.059$) GeV$^{-2}$ ,
$\alpha_0=-0.725\pm0.170$\\
${}^\dag$ fit without $K$: $\alpha=(0.846\pm0.013$) GeV$^{-2}$ , $\alpha_0=-0.431\pm0.042$
\end{table}

\begin{table}
  \caption{Fitted parameters of the $(n_r,M^2)$  Regge
    trajectories for light mesons.} 
  \label{tab:rtn}
\begin{ruledtabular}
\begin{tabular}{cccccc}
Meson&$\beta$ (GeV$^{-2}$)& $\beta_0$&Meson&$\beta$ (GeV$^{-2}$)& $\beta_0$\\
\hline
$q\bar q$&&& $s\bar s$\\
$\pi$& $0.679\pm0.023^*$& $-0.018\pm0.014^*$&$\eta_{s\bar s}$& $0.559\pm0.009$&$-0.315\pm0.026$\\
$\rho(^3S_1)$&$0.700\pm0.023$&$-0.451\pm0.060$ &$\varphi$& $0.597\pm0.009$&$-0.662\pm0.031$\\
$a_0$& $0.830\pm0.032$&$-1.214\pm0.109$ &$f_0$& $0.566\pm0.009$&$-1.156\pm0.039$\\
$a_1$& $0.840\pm0.037$&$-1.401\pm0.134$&$f_1$&$0.561\pm0.013$& $-1.224\pm0.058$\\
$b_1$& $0.863\pm0.030$&$-1.431\pm0.106$&$h_1$& $0.575\pm0.015$&$-1.292\pm0.066$\\
$a_2(^3P_2)$& $0.867\pm0.036$&$-1.585\pm0.134$&$f_2$& $0.581\pm0.007$&$-1.370\pm0.031$\\  
$\rho(^3D_1)$&$0.894\pm0.013$&$-2.182\pm0.050$ \\
$\pi_2$&$0.916\pm0.032$&$-2.514\pm0.134$ \\
$\rho_3(^3D_3)$&$0.874\pm0.041$&$-2.623\pm0.189$ \\
$a_2(^3F_2)$& $0.891\pm0.010$&$-2.881\pm0.043$\\
$a_3$& $0.890\pm0.014$&$-3.254\pm0.066$\\
$b_3$& $0.906\pm0.015$&$-3.225\pm0.071$\\
$a_4$& $0.899\pm0.016$&$-3.672\pm0.084$\\
\end{tabular}
 \end{ruledtabular}
${}^*$ fit without $\pi$: $\beta=(0.750\pm0.032)$ GeV$^{-2}$, $\beta_0=-0.287\pm0.109$
\end{table}

 It is
important to note that the quality of  fitting the $\pi$ meson Regge
trajectories both in ($J, M^2$) and $(n_r,M^2)$ planes is significantly
improved if the ground state $\pi$ is excluded from the fit (the
$\chi^2$ is reduced by more than an order of magnitude and becomes
compatible  with the values for other trajectories). 
In the kaon
case omitting the ground state also improves the fit but not so
dramatically as for the pion. 
The corresponding trajectories are shown in Figs.~\ref{fig:pir}, \ref{fig:kr} and
\ref{fig:pin} by dashed lines, the fitted values of the slopes and
intercepts are given in the footnotes to Tables~\ref{tab:rtj} and
\ref{tab:rtn}.  This indicates the special role of the pion
originating from the chiral symmetry breaking.

It can be seen in  Figs.~\ref{fig:rr} and \ref{fig:a0r} that
$\rho(1700)$ and $a_0(1450)$ do not lie on the corresponding Regge
trajectories. This further confirms our previous conclusion that
$a_0(1450)$ should be predominantly a tetraquark state and suggests
the possible exotic nature of $\rho(1700)$.

From the comparison of the slopes in Tables~\ref{tab:rtj},
\ref{tab:rtn} we see that the $\alpha$ values are systematically larger than
the $\beta$ ones. The ratio of their mean values is about 1.3 both for the light
$q\bar q$ isovector and $s\bar s$ mesons. Such ratio is lower than
predictions of the QCD string model (\ref{eq:str}) and massless
Salpeter equation (\ref{eq:slsr}). The mean value of the slope $\beta
\sim 0.85$~GeV$^{-2}$ for isovector mesons is about two times larger
than the result of the above mentioned models $\beta=1/(4\pi A)\approx
0.44$~GeV$^{-2}$, for $A=0.18$~GeV$^2$ used in our approach.     

The assignment of the experimentally observed states to the
corresponding Regge trajectories in our model based on their masses
and $J^{PC}$ values
(see Figs.~\ref{fig:rr}-\ref{fig:an}) is slightly different 
from the previous phenomenological analysis \cite{anisovich,afonin} based on
the equal values for the slopes $\alpha$ and $\beta$. However
the number of states, for which such correspondence is found, is
approximately the same. Future experimental data will shed 
further light on this issue.

\section{Conclusions}
\label{sec:concl}

The mass spectra of light quark-antiquark mesons were calculated in the
framework of the QCD-motivated
relativistic quark model. All considerations were
carried out without application of the unjustified nonrelativistic
$v/c$ expansion. Such approach leads to the nonlocal  quasipotential
of the relativistic quark-antiquark interaction. To make it local, the 
substitution  (\ref{eq:sub}) was used. As a result 
the relativistic local
quasipotential was obtained which depends on the mass of the meson in
a complicated nonlinear way. 
Such dependence allowed us to get masses of the $\pi$ and $K$ 
mesons in agreement with experimental data 
in the considered
model, where the chiral symmetry is explicitly broken by the
constituent quark masses.
It was found that the lightest scalar 
($1^3P_0$) states have masses above 1 GeV. This confirms our previous
conclusion \cite{ltetr} 
that $f_0(600)$ ($\sigma$), $K^*_0(800)$
($\kappa$), $f_0(980)$ and $a_0(980)$ could be the diquark-antidiquark
tetraquark states. It was found that the calculated masses of light
mesons reproduce the linear Regge trajectories both in  the
($J, M^2$) and
$(n_r,M^2)$ planes. Their slopes and intercepts were determined.  The
slope of the orbital excitations $\alpha$ was found to be
in average 1.3 times larger than the slope of the trajectories of
radial excitations $\beta$. This value of the ratio $\alpha/\beta$ is
smaller than the predictions based on the spinless Salpeter equation
(\ref{eq:slp}) and the QCD string model. The obtained slope $\beta$ for the
isovector light mesons is almost two times higher than the value predicted
by the above models. Possible experimental candidates
for the states populating the Regge trajectories are identified in
Figs.~\ref{fig:rr}-\ref{fig:an}. Predictions for the masses of the
missing states are presented in Tables~\ref{tab:nsmm},\ref{tab:smm}. 
It is clearly seen that the chiral symmetry is not 
restored for highly excited states in our model. This should be 
expected since the Lorentz-scalar part of the confining potential
explicitly breaks the chiral symmetry.   
Our results in some cases differ from the previous 
phenomenological prescriptions \cite{anisovich,afonin}. Future experimental
data  can help in discriminating between the theoretical predictions.

The authors are grateful to S. Gerasimov, H. Forkel, V. Matveev,
V. Savrin, D. Shirkov and A. Vainshtein for support and discussions.  
This work was supported in part by
the Russian Science Support Foundation 
(V.O.G.) and the Russian Foundation for Basic Research (RFBR), grant
No.08-02-00582 (R.N.F. and V.O.G.).


\begin{thebibliography}{99}
\bibitem{bugg} D. V. Bugg,  Phys. Rep. {\bf
    397}, 257 (2004).
\bibitem{Anisovich}
  A.~V.~Anisovich {\it et al.},
  Phys.\ Lett.\  B {\bf 542}, 19 (2002).
\bibitem{at}
  C.~Amsler and N.~A.~Tornqvist,
  Phys.\ Rep.\  {\bf 389}, 61 (2004).

\bibitem{kz}
  E.~Klempt and A.~Zaitsev,
  Phys.\ Rep.\  {\bf 454}, 1 (2007).

\bibitem{glb}
  V.~Crede and C.~A.~Meyer,
  arXiv:0812.0600 [hep-ex].

\bibitem{achasov}
  N.~N.~Achasov,
  arXiv:0810.2601 [hep-ph].





\bibitem{glozman} L. Ya. Glozman, 
Phys. Rep. {\bf 444}, 1 (2007).
\bibitem{shifvain} M.~Shifman and A.~Vainshtein,
  Phys.\ Rev.\  D {\bf 77}, 034002 (2008).

\bibitem{gi} S. Godfrey and N. Isgur, { Phys. Rev. D} {\bf 32}, 189
  (1985). 
\bibitem{mr} P. Maris and C.D. Roberts, { Int.J. Mod. Phys. E} {\bf
    12}, 297 (2003); P. Maris and P.C. Tandy, { Phys. Rev. C} {\bf
    60}, 055214 (1999).
\bibitem{k} M.~Koll, R.~Ricken, D.~Merten, B.~C.~Metsch and H.~R.~Petry,
  Eur.\ Phys.\ J.\  A {\bf 9}, 73 (2000). 
\bibitem{bnps}
  M.~Baldicchi, A.~V.~Nesterenko, G.~M.~Prosperi, D.~V.~Shirkov and C.~Simolo,
  Phys.\ Rev.\ Lett.\  {\bf 99}, 242001 (2007).
\bibitem{lse}
   N.~Ligterink and E.~S.~Swanson,
   Phys.\ Rev.\  C {\bf 69}, 025204 (2004); F.~J.~Llanes-Estrada and
   S.~R.~Cotanch, 
   Nucl.\ Phys.\  A {\bf 697}, 303 (2002).
\bibitem{erv} D. Ebert, H. Reinhardt and M.K. Volkov, {
    Prog. Part. Nucl. Phys.} {\bf 33}, 1 (1994); M.K. Volkov, D. Ebert
    and M. Nagy, { Int. J. Mod. Phys. A} {\bf 13}, 5443
    (1998); M.K. Volkov and C. Weiss,  Phys. Rev. D {\bf 56}, 221 (1997).
\bibitem{kp}  N.V. Krasnikov and A.A. Pivovarov, Phys. Lett. B {\bf
   112}, 397 (1982); 
N.V. Krasnikov, A.A. Pivovarov and N.N. Tavkhelidze,  Z. Phys. C {\bf
   19}, 301 (1983);  A.L. Kataev, hep-ph/9805218.
\bibitem{ak} A. Ali Khan et al., Phys. Rev. D {\bf 65}, 054505 (2002).
\bibitem{fork}
  H.~Forkel, M.~Beyer and T.~Frederico,
  JHEP {\bf 0707}, 077 (2007).
\bibitem{af} S. S. Afonin, arXiv:0903.0322 [hep-ph].
\bibitem{lmes}
  D.~Ebert, R.~N.~Faustov and V.~O.~Galkin,
  Mod.\ Phys.\ Lett.\  A {\bf 20}, 1887 (2005).
\bibitem{lmesff}
  D.~Ebert, R.~N.~Faustov and V.~O.~Galkin,
  Eur.\ Phys.\ J.\  C {\bf 47}, 745 (2006).
\bibitem{ltetr}
  D.~Ebert, R.~N.~Faustov and V.~O.~Galkin,
Eur. Phys. J. C {\bf 60}, 273 (2009).

\bibitem{anisovich}
  A.~V.~Anisovich, V.~V.~Anisovich and A.~V.~Sarantsev,
  Phys.\ Rev.\  D {\bf 62}, 051502 (2000);
V.~V.~Anisovich, L.~G.~Dakhno, M.~A.~Matveev, V.~A.~Nikonov and A.~V.~Sarantsev,
  Phys.\ Atom.\ Nucl.\  {\bf 70}, 450 (2007).


\bibitem{efg} D. Ebert, R.N. Faustov and V.O. Galkin, {
    Phys. Rev. D} {\bf 67}, 014027 (2003).
\bibitem{egf} D. Ebert, V.O. Galkin and R.N. Faustov, {
    Phys. Rev. D} {\bf 57}, 5663 (1998).



\bibitem{afonin}
  S.~S.~Afonin,
  Phys.\ Rev.\  C {\bf 76}, 015202 (2007)
\bibitem{bs} H.A. Bethe and E.E. Salpeter, {\it Quantum Mechanics of
    One-and Two-Electron Atoms} (Springer-Verlag, Berlin, 1957).
\bibitem{bvb} A.M. Badalian, A.I. Veselov and B.L.G. Bakker, {
    Phys. Rev. D} {\bf 70}, 016007 (2004); Yu.A. Simonov, {
    Phys. Atom. Nucl.} {\bf 58}, 107 (1995).
\bibitem{shirkov}
D.~Shirkov,
  arXiv:0807.1404 [hep-ph];
  D.~V.~Shirkov and I.~L.~Solovtsov,
  Phys.\ Rev.\ Lett.\  {\bf 79}, 1209 (1997).
\bibitem{pdg} C. Amsler {\it et al.} (Particle Data Group),
  Phys. Lett. B667, 1 (2008). 

\bibitem{fks}
  T.~Feldmann, P.~Kroll and B.~Stech,
  Phys.\ Rev.\  D {\bf 58}, 114006 (1998).



\bibitem{schechter}
  D.~Black, A.~H.~Fariborz and J.~Schechter,
  Phys.\ Rev.\  D {\bf 61}, 074001 (2000);  A.~H.~Fariborz, R.~Jora and J.~Schechter,
  arXiv:0902.2825 [hep-ph].



\bibitem{maiani}
 G.~'t~Hooft,
  G.~Isidori, L.~Maiani, A.~D.~Polosa and V.~Riquer, 
Phys.\ Lett.\  B {\bf 662}, 424 (2008)

\bibitem{simonov}
  Yu.~S.~Kalashnikova, A.~V.~Nefediev and Yu.~A.~Simonov,
  Phys.\ Rev.\  D {\bf 64}, 014037 (2001); A.~Y.~Dubin, A.~B.~Kaidalov
  and Yu.~A.~Simonov, 
  Phys.\ Lett.\  B {\bf 323}, 41 (1994); T.~J.~Allen, C.~Goebel,
  M.~G.~Olsson and S.~Veseli, 
  Phys.\ Rev.\  D {\bf 64}, 094011 (2001).



\bibitem{bicudo}
  P.~Bicudo,
  Phys.\ Rev.\  D {\bf 76}, 094005 (2007).




\end{thebibliography}
\end{document}